\documentclass[twocolumn,showpacs,amsmath,amssymb,prb,superscriptaddress,a4paper,floatfix,showkeys]{revtex4}
\usepackage{dcolumn}
\usepackage{graphicx}
\usepackage{epsfig}

\begin{document}

\title{Event-by-event simulation of a quantum eraser experiment\footnote{Accepted for publication in J. Comput. Theor. Nanosci.}}

\author{F. Jin}
\affiliation{Department of Applied Physics, Zernike Institute for Advanced Materials,
University of Groningen, Nijenborgh 4, NL-9747 AG Groningen, The Netherlands}

\author{S. Zhao}
\affiliation{Department of Applied Physics, Zernike Institute for Advanced Materials,
University of Groningen, Nijenborgh 4, NL-9747 AG Groningen, The Netherlands}

\author{S. Yuan}
\affiliation{
Institute of Molecules and Materials, Radboud University of Nijmegen,
NL-6525ED Nijmegen, The Netherlands}

\author{H. De Raedt}

\email{h.a.de.raedt@rug.nl}

\affiliation{Department of Applied Physics, Zernike Institute for Advanced Materials,
University of Groningen, Nijenborgh 4, NL-9747 AG Groningen, The Netherlands}

\author{K. Michielsen}

\affiliation{
Institute for Advanced Simulation, J\"ulich Supercomputing Centre, Research Centre Juelich, D-52425 Juelich, Germany
}


\date{\today}

\begin{abstract}
We present a computer simulation model that is a one-to-one copy of
a quantum eraser experiment with photons (P. D. D. Schwindt {\sl et al.}, Phys. Rev. A 60, 4285 (1999)).
The model is solely based on experimental facts,
satisfies Einstein's criterion of local causality and does not require knowledge of
the solution of a wave equation.
Nevertheless, the simulation model reproduces the averages as obtained from
the wave mechanical description of the quantum eraser experiment,
proving that it is possible to give a particle-only description of
quantum eraser experiments with photons.
We demonstrate that although the visibility can be used as a measure for the interference, it cannot
be used to quantify the wave character of a photon. The classical particle-like simulation model
renders the concept of wave-particle duality, used to explain the outcome of the quantum eraser experiment
with photons, superfluous.
\end{abstract}

\maketitle

\section{Introduction}

According to wave-particle duality, a concept of quantum theory (QT), photons
exhibit both wave and particle behavior depending upon the circumstances of
the experiment~\cite{HOME97}.
The wave and particle behavior of photons is believed to be complementary.
When we know (observe) the which-way (WW) information (particle behavior),
there is no interference pattern (wave behavior)~\cite{FEYN65}.
Parameters quantifying the interference and the WW information are the visibility  ${\cal V}$ and
the path distinguishability ${\cal D}$, respectively.
According to the complementarity relation of QT, ${\cal V}^2 + {\cal D}^2 \leq 1$~\cite{JAEG95,ENGL96}.

In 1982, Scully and Dr{\"{u}}hl proposed a photon interference experiment,
called ``quantum eraser''~\cite{SCUL82}, in which
the photons are labelled by WW markers (three-level atoms).
In this experiment, we know (but not observe)
the WW information of the photons and then we expect that there is no interference.
However by erasing the WW information afterwards by a ``quantum eraser'', the interference pattern
can be recovered~\cite{SCUL82}.
The interference pattern can even be recovered after the data have already
been recorded and saved in a file~\cite{KIM00}.

Quantum eraser experiments have been described ``as one of the most intriguing effects in quantum mechanics'',
but have also been regarded as ``the fallacy of delayed choice and quantum eraser''~\cite{AHAR05}.
Clearly, they challenge the point of view
that the wave and particle behavior of photons are complementary:
The observation of interference, commonly associated with wave behavior,
depends on the way the data is analyzed after the photons
have passed through the interferometer.

The question that we answer in the affirmative in this paper is: ``Can we
simulate a quantum eraser experiment without invoking concepts of quantum
theory and without first solving the wave mechanical problem?''

\begin{figure}[t]
\begin{center}
\includegraphics[width=8cm]{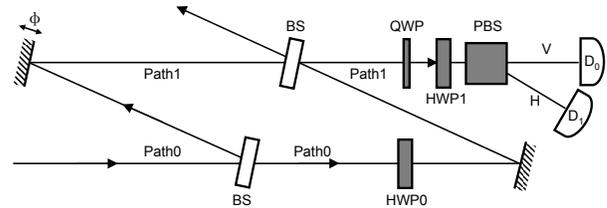}
\caption{Schematic diagram of the experimental setup for the quantum eraser experiment
with photons studied in Ref.~\onlinecite{SCHW99}.
BS: beam splitter; PBS: polarizing beam splitter;
HWP0 and HWP1: half-wave plates; QWP: quarter-wave plate; $D_0$, $D_1$: detectors;
$\phi$: phase shift introduced in Path1.
}
\label{eraser}
\end{center}
\end{figure}

\subsection{Quantum eraser with photons}
\subsubsection{Experimental realization}
The quantum eraser has been implemented in several different experiments
with photons, atoms, etc.~\cite{SCUL91,KWIA92,PITT96,SCHW99,KIM00,WALB02,SCAR07}.
Although much more difficult to realize experimentally, quantum erasers
may also be realized with quantum dots~\cite{HACK98,KANG07} and mesoscopic electromechanical
devices~\cite{ARMO01}.

In Ref.~\cite{SCHW99}, Schwindt {\sl et al.}
reported an experimental realization of a quantum eraser
in which the polarization of the photons has been used to encode the WW information.
In this paper, we focus on this particular experiment.
The experimental setup (see Fig.~\ref{eraser}) consists of a linearly polarized
single-photon source (not shown), a Mach-Zehnder interferometer (MZI)
of which the length of Path1 (see Fig.~\ref{eraser})
can be varied, inducing a relative phase shift $\phi$ between
Path0 and Path1,
an adjustable analysis system which is a combination of a quarter-wave plate (QWP),
a half-wave plate (HWP) HWP1, and a calcite prism operating as a polarizing beam splitter (PBS).
Another adjustable HWP, HWP0, is inserted in Path0 of the MZI
to entangle the photon's path with its polarization.

According to Ref.~\cite{SCHW99} the pictorial description of the experimental observations is as follows.
If a photon, described by a pure, vertically polarized state $V$ is injected into the interferometer with the HWP0 set to $45^\circ$,
then the photon that arrives at the second beam splitter (BS) of the MZI carries a WW marker:
The photon is in the horizontally polarized state $H$ if it followed Path0 and it
is in the $V$ state if it followed Path1.
If the optical angle of HWP1 is zero,
there will be no interference (${\cal V}=0$) and the detectors give us the full WW information of each detected photon (${\cal D}=1$).
If the optical angle of HWP1 is nonzero, the $H$ and $V$ states interfere ($0<{\cal V}\leq 1$) and
the WW information of each photon will be partially or completely ``erased'' ($0\leq {\cal D}<1$).
Thus, by varying the optical angle $\theta_{HWP1}$ of HWP1, the illusion is created
that the character of the photon in the MZI ``changes'' from particle
to wave and vice versa.
If photons described by a completely mixed, that is an unpolarized, state are emitted, then no WW information can be obtained and
also no interference can be observed (${\cal D}={\cal V}=0$), independent of the orientation of HWP0.
However, varying $\theta_{HWP1}$ can still lead to a recovery of interference ($0< {\cal V}\leq 1$).
For photons described by a partially mixed state, a state that can we expressed as containing a completely mixed component and a pure component,
partial WW information can be obtained.
Since the completely mixed component contains no WW information and displays no interference, the maxima of ${\cal D}$ and ${\cal V}$
are smaller than one and numerical equal to the state purity.
Also in this case complete visibility can be recovered by varying $\theta_{HWP1}$.

\subsubsection{Event-by-event simulation model}
It is important to realize that the counter-intuitive
features of quantum eraser experiments result from attempts
to apply the concepts and the formalism of QT to
a description of the experimental results in terms of individual events~\cite{HOME97}.
Logically speaking, there are two possibilities:
\begin{enumerate}
\item{We accept the postulate that it is fundamentally impossible to give a logically
consistent description of the experimental results in terms of individual events, that is we accept
that there is no explanation that goes beyond the quantum theoretical description in terms of averages
over many events.}
\item{We search for an explanation of the experimental facts that
goes beyond a description in terms of averages.}
\end{enumerate}
In this paper, we demonstrate that the second option is a viable one.
Thus, we adopt the point of view that although
QT correctly predicts averages of many detection events, it has
nothing to say about individual events~\cite{HOME97}.

We propose an event-by-event simulation model
that is a one-to-one copy of the quantum eraser experiment reported in Ref.~\cite{SCHW99}.
The simulation model describes a particle-like, classical, local and causal dynamical system.
Each component of the laboratory experiment such as the single-photon source,
the BS, HWP, QWP, and PBS are simulated by corresponding algorithms. 
By connecting the output(s) of one component
to the input(s) of another one, we construct the simulation equivalent
of the experimental setup depicted in Fig.~\ref{eraser}.
By construction this network of dynamical systems
satisfies Einstein's criterion of local causality.
The data is analyzed by counting the detection events, just as in the real experiment.

We demonstrate that our model reproduces the results of QT, that is
the averages predicted by QT and confirmed by experiment~\cite{SCHW99},
without first solving a wave equation. In fact, we show that
it is possible to give an entirely classical, particle-only description
for the single-photon quantum eraser experiment reported in Ref.~\cite{SCHW99}.
We show that the interference patterns, commonly associated with wave behavior, can be built up by many
particles having full WW information (we can always track the photons during the simulation) that arrive
one-by-one at a detector.

The work of this paper builds on earlier work~\cite{RAED05d,RAED05b,RAED05c,MICH05,RAED06c,RAED07a,RAED07b,RAED07c,ZHAO07b,ZHAO08,ZHAO08b,JIN09a,JIN09b}
that demonstrates that quantum phenomena can be simulated on the level of individual events
without first solving a wave equation and even without invoking concepts of QT, wave theory or probability theory.
Specifically, in our earlier work we have demonstrated that it is possible to simulate event-by-event, a
single-photon beam splitter and Mach-Zehnder interferometer experiments, 
Einstein-Podolsky-Rosen-Bohm experiments with photons, 
Wheeler's delayed choice experiment with single photons,
the double-slit and two-beam interference, 
quantum cryptography protocols,
and universal quantum computation. 
The latter proves that in principle we can perform an event-by-event (particle-like) simulation of any quantum system~\cite{NIEL00}.
Some interactive demonstration programs are available for download~\cite{COMPPHYS,MZI08,DS08}.

\subsection{Irrelevance of Bell's theorem}
It is not uncommon to find in the recent literature,
statements that it is impossible to simulate quantum phenomena by classical processes.
Such statements are thought to be a direct consequence of Bell's theorem~\cite{BELL93} but
are in conflict with other work that has pointed out the irrelevance of Bell's theorem
~\cite{PENA72,FINE74,FINE82,FINE82a,FINE82b,MUYN86,JAYN89,BROD93,FINE96,KHRE99,SICA99,BAER99,%
HESS01,HESS05,ACCA05,KRAC05,SANT05,MORG06,KHRE07,ADEN07,NIEU09,MATZ09,RAED09a}.
A survey of the literature suggests that, roughly speaking,
physicists can be classified as those who believe in the reasonableness of Bell's arguments,
those who advance logical and mathematical arguments to show that a violation
of Bell's (and related) inequalities does not support the far-reaching
conclusions of the former group of physicists
and those who do not care about Bell's theorem at all.
The authors of this article belong to the second group.

Although we expect discussions of philosophical or metaphysical
aspects of Bell's theorem to continue forever,
as explained in a review article that has appeared in this journal~\cite{RAED07c},
from the viewpoint of simulating quantum phenomena on a digital computer,
Bell's no-go theorem is of no relevance whatsoever.

This conclusion is supported by several explicit examples that prove
that it is possible to construct algorithms that satisfy
Einstein's criteria for locality and causality, yet reproduce
{\sl exactly} the two-particle correlations of a quantum system in the singlet state,
without invoking any concept of quantum theory~\cite{RAED06c,RAED07a,RAED07b,RAED07c,RAED07d,ZHAO08}.
It is therefore an established fact
that purely classical processes can produce the correlations
that are characteristic for a quantum system in an entangled state,
thereby disposing of the mysticism that is created by Bell's no-go theorem.

The key point is to realize that QT or the probabilistic models
proposed by Bell cannot, on a fundamental level, address the (non)existence
of algorithms, that is of well-defined processes,
that give rise to the distributions of the events,
described by these theories/models.

The philosophy behind our simulation approach is very simple:
If we can construct an algorithm that
\begin{enumerate}
\item{does not rely on the solution of a wave  equation,}
\item{satisfies the elementary
criteria of locality and causality as formulated by Einstein,}
\item{produces data of the same type as
the data collected in the laboratory experiment,}
\item{by analyzing the simulated data according
to the procedure used to analyze the experimental data
leads to the same conclusion, namely that
certain averages of the raw data agree with the
quantum theoretical description of the whole experiment,}
\item{contains algorithms that simulate the various components
(beam splitter, etc.) of the experiment and can, with no change,
be re-used to simulate other experiments,}
\end{enumerate}
then we may conclude that we have built a simulation model
for the laboratory experiment.

Loosely speaking, if the experimenter would be unable to
distinguish between data recorded in a genuine
experiment and data provided by the simulation algorithm,
then the experiment has been
``de-mystified'' in the sense that we have found
a process that offers a description of the observed
phenomena on the level of individual events
and without invoking (concepts of) wave theory.

To avoid possible misunderstandings,
the work presented in this paper is not concerned with an
interpretation or an extension of QT
nor does it affect the validity of QT as such.
QT describes the collective result of many events,
that is averages of many events, extremely well but does not provide
a description on the level of individual clicks of a detector~\cite{HOME97}.

\subsection {Structure of the paper}
Section~\ref{sec2} reviews the standard concepts of QT that are needed
to give a quantum theoretical treatment of the quantum eraser experiment~\cite{SCHW99}.
Section~\ref{sec3} discusses the general ideas that underpin our event-by-event simulation
approach. We address the fundamental problem of reconciling the observation
of ``clicks'' with a wave mechanical theory from the viewpoint of algorithms,
processes and computation.
We show that in general, it is impossible to attribute ``clicks'' to individual
wave amplitudes and explain how our simulation approach circumvents this
fundamental problem.
Section~\ref{sec4} explains how the pure and mixed states
of a quantum systems can be represented in our simulation approach.
In Section~\ref{sec5}, we specify the simulation model in full detail.
Data of event-by-event simulations of the quantum eraser experiment
are presented in Section~\ref{sec6}. We show that our classical, particle-like
simulation model reproduces all the results of QT for this experiment.
Our conclusions can be found in Section~\ref{sec7}.

\section{Quantum theory}\label{sec2}

In QT, a system is described by the state $\left|\alpha \right>$, a vector in a Hilbert space~\cite{BALL03}.
This vector can be written as a linear combination of a
complete set of orthonormal basis states $\left|i\right>$ for $i=1,\dots,d$ where
$d$ denotes the dimension of the Hilbert space.
These basis states are chosen such that they facilitate the formulation of the model.
The amplitude for a quantum system to go from a state
$\left|\alpha \right>$ to another state $\left|\beta \right>$
is given by $\left<\beta |\alpha \right> = \sum_{i=1}^d \left<\beta |i \right>\left<i|\alpha \right>$.
With respect to the basis states $\{\left |i\right>\}$,
the optical apparatus $T$ is defined through its transition matrix elements $\left<i|T|j\right>$.
If the optical apparatus $T$ induces a transition from the state
$\left|\alpha \right>$ to the state $\left|\beta \right>$,
the amplitude for this transition
is given by $\left<\beta |T| \alpha \right> = \sum_{i,j=1}^d \left<\beta |i\right>\left<i|T|j\right>\left<i|\alpha \right>$.
Finally, the probability $\mathrm{Prob}(\beta,\alpha)$ for this transition to occur is related to the amplitude through the Born rule
\begin{equation}
\mathrm{Prob}(\beta,\alpha) = \vert \left<\beta |T| \alpha \right> \vert^2.
\end{equation}

According to the above scheme, we can easily calculate the predictions of QT for the experiment shown in Fig.~\ref{eraser}.
The basis states correspond to $H$ or $V$ polarized photons that travel along Path0 or Path1.
The transition matrices of the optical components such as the BS, PBS, HWP and QWP can
be found in Ref.~\cite{ENGL01} and in the appendix.
In the appendix, we also give the quantum theoretical expressions for the
visibility for the experiment depicted in Fig.~\ref{eraser}
that will be used for the comparison with our simulation results.

The above formulation assumes that the quantum system is in the pure state~\cite{BALL03}.
Some of the experiments reported in Ref.~\cite{SCHW99} require a description in terms of a mixed state~\cite{BALL03}.
A system is in a mixed state if it is in one of its $m$ pure states
$|{\alpha_1} \rangle,|{\alpha_2}\rangle,\cdots,|{\alpha_m}\rangle$
with probability $p_{1},p_{2},\cdots,p_{m}$, respectively~\cite{BALL03}.
A quantum system in a mixed state is conveniently
described through the density matrix~\cite{BALL03}
\begin{equation}
\rho = \sum_{j=1}^m 	 p_{j}|\alpha_j\rangle \langle \alpha_j|,
\label{mixedstate}
\end{equation}
where it is assumed that
$\sum_{j=1}^m p_j = 1$, $p_j\geq 0$ for $j=1,\ldots,m$,
and that the states
$|\alpha_j\rangle$ are normalized such that $\mathbf{Tr} \rho=1$.
According to QT, for a system in a mixed state $\rho$, the expectation value of the operator $\Omega$ is given by~\cite{BALL03}
\begin{equation}
\left\langle \Omega\right\rangle = \mathbf{Tr} \rho \Omega = \sum_{j=1}^m p_j
\langle \alpha_j | \Omega |\alpha_j\rangle .
\label{eq_mxs}
\end{equation}

\section{Event-by-event simulation}\label{sec3}

Our event-based simulation approach is unconventional in that it does not require
knowledge of the wave amplitudes obtained by first solving the quantum theoretical problem
nor do we first calculate the quantum potential (which
requires the solution of the Schr{\"o}dinger equation) and
then compute the Bohm trajectories of the particles.
Instead, the detector clicks are generated event-by-event by locally causal,
adaptive, classical dynamical systems.
Our approach employs algorithms, that is we define processes, that contain
a detailed specification of each individual event
which, as we now show, cannot be derived from a wave theory such as QT.

To understand the subtleties that are involved, it is helpful to consider
a simple example.
Let us consider the MZI unit of the quantum eraser
and omit the polarization label of the photons.
According to QT, the amplitudes $b_0$ and $b_1$ to observe a photon
in Path0 or Path1 after the second BS are related to the input amplitudes
$a_0$ and $a_1$ by~\cite{RARI97}
\begin{eqnarray}
\left(
\begin{array}{c}
b_0\\
b_1
\end{array}
\right)
&=&
\frac{1}{2}
\left(
\begin{array}{cc}
1&i\\
i&1
\end{array}
\right)
\left(
\begin{array}{cc}
e^{i\phi_0}&0\\
0&e^{i\phi_1}
\end{array}
\right)
\left(
\begin{array}{cc}
1&i\\
i&1
\end{array}
\right)
\left(
\begin{array}{c}
a_0\\
a_1
\end{array}
\right)
,\cr
&\equiv & ABA
\left(
\begin{array}{c}
a_0\\
a_1
\end{array}
\right).
\label{MZ1}
\end{eqnarray}
Let us assume that $a_0=1$ and $a_1=0$, meaning that the photons
enter the MZI through Path0 only.
The probabilities $P_0$ ($P_1$) for a click in detector $D_0$ ($D_1$) are
given by
\begin{equation}
P_k
=
\left|
\sum_{j=0,1}
\sum_{i=0,1}
A_{k,j}B_{j,i}A_{i,0}
\right|^2
\quad,\quad k=0,1.
\label{MZ2}
\end{equation}

Using Eqs.~(\ref{MZ1}) and (\ref{MZ2}) a simple calculation yields a closed form
expression for $P_k$. Once we know $P_k$, it is trivial to use it as input for a process that
generates clicks of the detectors $D_0$ and $D_1$.
This approach relies on what we call the ``solution'' of the quantum theoretical problem.
It is irrelevant whether we have a closed form expression for $P_k$ or only know
$P_k$ in tabulated form. The point is that we analytically worked out the
sums over the indices $i$ and $j$ in Eq.~(\ref{MZ2}).
Let us now assume that we do not know how to perform the
sums over the indices $i$ and $j$ in Eq.~(\ref{MZ2}) by ourselves
and that there is some ``magical process'' that carries out
the sum for us. In other words, we assume that we do not know
$P_0$ and $P_1$.

In practice, any process that performs the sums
in Eq.~(\ref{MZ2}) by selecting (one-by-one) the pairs $(i,j)$ from the set
${\cal S}={(0,0),(1,0),(0,1),(1,1)}$ defines a sequence of ``events'' $(i,j)$.
The key question now is: Can we identify the selection
of the pairs with ``clicks'', events registered by
a detector? We now prove that this is impossible.

A characteristic feature of all wave phenomena
is that not all contributions to the sums
in Eq.~(\ref{MZ2}) have the same sign:
In wave theory, this feature is essential
to account for destructive interference.
But, at the same time this feature
forbids the existence of a process of which the ``events'' can
be identified with the clicks of the detector.

This is easily seen by considering a situation
for which, for instance, $P_0=0$.
In this case, the detector $D_0$ should never click.
However, according to Eq.~(\ref{MZ2}), the process that samples
from the set ${\cal S}$ produces ``events''
such that the sums over all these ``events'' vanishes.
Therefore, if we want to identify these ``events''
with the clicks that we observe, we run into a logical contradiction:
To perform the sums in Eq.~(\ref{MZ2}), we have to generate events
that in the end cannot be interpreted as clicks since in this
particular case no detector clicks are observed.

Thus, the conclusion is that the individual terms in expression
Eq.~(\ref{MZ2}) do not contain the ingredients to define a process that generates
the clicks of the detectors that we observe.

The crux of our event-by-event simulation approach is that we do not start
from expression Eq.~(\ref{MZ2}) but construct
a process that converges to Eq.~(\ref{MZ2})
while generating events that correspond to
the observed events.
To grasp this idea, consider
the well-known Metropolis Monte Carlo (MMC) method
for solving statistical mechanical problems~\cite{HAMM64,LAND00}.
The MMC method generates states $S$, events in our terminology,
with a probability density~\cite{HAMM64,LAND00}
\begin{equation}
P(S)=\frac{e^{- E(S)/k_BT}}{\sum_S e^{-E(S)/k_BT}}
,
\label{sub2}
\end{equation}
where $E(S)$ denotes the energy of the state $S$,
$k_B$ is Boltzmann's constant and $T$ is the temperature.
At first sight, sampling from Eq.~(\ref{sub2}) is impossible
because in all but a few nontrivial cases for which the
partition function $\sum_S e^{-E(S)/k_BT}$ is known,
we do not know the denominator.
MMC solves this problem by constructing a Markov chain that generates
a sequence of events $S$ such that asymptotically
these events are distributed according
to the (unknown) probability density Eq.~(\ref{sub2})~\cite{HAMM64,LAND00}.

The analogy with our event-by-event simulation approach is the following.
Although very different in all technical details,
our event-based method uses a deterministic process
of which the sampling distribution
converges to the unknown (by assumption) probability distribution $P_k$ for $k=0,1$.
Initially, the system does not know about this limiting
probability distribution and hence, during 
a short transient period, the frequencies with which
events are generated may not correspond to this distribution.
However, for many events, which is the situation
described by QT, these first few ``wrong'' events
disappear in the statistical fluctuations and are therefore irrelevant
for the comparison of our event-based simulation results with QT.
It should be clear that the foregoing
does not depend on the specific example that we used
for the purpose of illustration.

Let us now discuss the general aspects of our simulation approach.
The simulation algorithms that we construct are
most easily formulated in terms of events, messages,
and units that process these events and messages.
Taking the quantum eraser experiment as an example,
in a pictorial description, the photon is regarded as a messenger,
carrying a message that represents its time-of-flight (phase) and polarization.
In this pictorial description, we may speak of ``photons'' generating
the detection events. However, these so-called photons, as we will call them in the following,
are elements of a model or theory for the real laboratory experiment only.
The only experimental facts are the settings of the various apparatuses and the detection events.
What happens in between activating the source and the registration of the detection
events belongs to the domain of imagination.

The processing units mimic the role of the optical components in the experiment
and the network by connecting the processing units represents the complete experimental setup.
The standard processing units consist of an input stage,
a transformation stage and an output stage.
The input (output) stage may have several channels
at (through) which messengers arrive (leave).
Other processing units are simpler in the sense that the input stage
is not necessary for the proper functioning of the device.
A message is represented by a set of numbers, conventionally represented by a vector.
As a messenger arrives at an input channel of a processing unit,
the input stage updates its internal state, represented by a vector, and
sends the message together with its internal state
to the transformation stage that implements the operation of the particular device.
Then, a new message is sent to the output stage
which selects the output channel through which the messenger will leave the unit.
At any given time, there is only one messenger being routed through the whole network.
There is no direct communication between the messengers.
From this general description, it should already be clear that the process that
is generated by the collective of classical dynamical systems is locally
causal in Einstein's sense.
Our simulation approach does not rely on concepts of probability theory
but instead, it generates events by way of classical, dynamical processes,
the frequencies of events of which converge to the quantum theoretical results
as the dynamical system relaxes to its stationary state.

\section{Simulation of pure and mixed states}\label{sec4}
In QT, the pure state is a description of the whole experiment, not of the individual events that
are recorded by the detectors~\cite{HOME97,BALL03}.
In our simulation approach, the messages carried by the messengers represent the pure state,
corresponding to a density matrix of the form
$\rho=\left|\alpha_k\right\rangle \left\langle \alpha_k \right|$,
that is $p_j=0$ for all $j\not=k$ and $p_k=1$.
In our simulation approach, the messages are constructed such that a large collection
of them yields the same averages as those we obtain from quantum theory.
Loosely speaking, we may say that a set of $N$ ($N$ sufficiently large)
messages of a certain type correspond to a pure state.

In the more general case, QT describes the whole experiment through the mixed state Eq.~(\ref{mixedstate}).
We simulate the mixed state by the following procedure.
Given $p_1,\ldots,p_m$, we pick an index $k\in\{1,\ldots,m\}$ using a pseudo-random number
and then send $N_k$ messages of type $k$ (corresponding to the pure state $|\alpha_k\rangle$) through the
network of processing units that represent the quantum system.
The precise value of $N_k$ is unimportant, as long at it is large enough to let the classical
dynamical system mimic the pure state $|\alpha_k\rangle$.

For the case at hand, the quantum eraser, the source can emit a pure state, a linear combination
of $V$ and $H$ polarized photons, or it can produce a mixed state of the two~\cite{SCHW99}.
Thus, we have $m=2$ and it what follows we will label the $N$'s by the subscripts $V$ and $H$
to facilitate the comparison with the terminology used in the experiment~\cite{SCHW99}.
Although not essential, in our simulation we simply choose $N_V=N_H$ and denote
the probabilities for the $V$- and $H$-polarized photons in a mixed state
by $p_V$ and $p_H$, respectively.

\section{Simulation model}\label{sec5}

As explained earlier, our simulation algorithm can be viewed as a message-processing
and message-passing process: It routes messengers, representing the photons, through a network of
message-processing units, playing the role of the optical components in the laboratory experiment.
In what follows we give a detailed description of each of the components of the network
representing the complete experimental setup of the quantum eraser experiment,
schematically depicted in Fig.~\ref{eraser}.

\begin{figure}[t]
\begin{center}
\includegraphics[width=8cm]{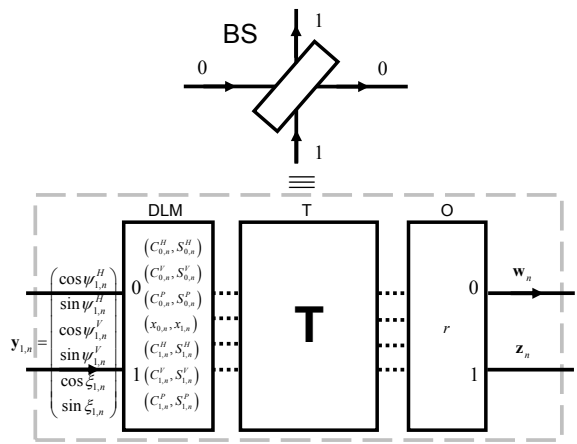}
\caption{Diagram of a DLM-based processing unit that performs an event-based simulation of a
beam splitter (BS).
The processing unit consists of three stages: An input stage (DLM), a transformation stage (T) and an output stage (O).
The solid lines represent the
input and output channels of the BS. The presence of a message is indicated
by an arrow on the corresponding channel line. The dashed lines indicate the
data flow within the BS. The transformation matrix ${\bf T}$ is given in Eq.~(\ref{T_vector}).}
\label{BS}
\end{center}
\end{figure}

\subsection{Messenger}
A messenger has its own internal clock, the hand of which rotates with frequency $f$.
When the messenger is created, the hand of the clock is set to time zero.
As the messenger travels from one position in space to another, the clock encodes the time-of-flight modulo the period $1/f$.
The message, the position of the clock's hand, is most conveniently represented by a two-dimensional unit vector
${\bf e}_l = (e_{0,l},e_{1,l})=(\cos \psi_l, \sin \psi_l)$, where $\psi_l = 2 \pi f t$, the subscript $l\geq0$ labeling the successive messages.
The messenger travels with a speed $c/n$ where $n$ is the refractive index of the medium in which the messenger moves and $c$ is the light velocity.
Clearly, this messenger is the event-based equivalent of a classical, linearly polarized electromagnetic wave with frequency $f$: The messenger corresponds to the light ray with wave vector ${\bf k}(k=2\pi f/c)$ and the clock mimics one of the electric field components in the plane orthogonal to ${\bf k}$~\cite{BORN64}.
Adding another clock to the messenger suffices to model the second electric field component orthogonal to the first one, and hence the fully polarized wave~\cite{ZHAO08b}.

Thus, each messenger carries a message represented by a
six-dimensional unit vector
\begin{equation}
	{\bf y}_{k,l}=\left(\begin{array}{c} \cos \psi_{k,l}^{H} \\
	 \sin \psi_{k,l}^{H} \\
	 \cos \psi_{k,l}^{V} \\
	 \sin \psi_{k,l}^{V} \\
	 \cos \xi _{k,l} \\
	 \sin \xi _{k,l}\end{array}\right).
\label{message}
\end{equation}
where the superscript $H$ ($V$) refers to the horizontal (vertical)
component of the polarization
and $\psi _{k,l}^{H}$, $\psi _{k,l}^{V}$, and $\xi _{k,l}$
represent the time of flight and polarization of the photon, respectively.
It is evident that the representation used here maps one-to-one
to the plane-wave description of a classical electromagnetic field~\cite{BORN64},
except that we assign these properties to each individual messenger, not to a wave.
The subscript $l\geq 0$ numbers
the consecutive messages and $k=0,1$ labels the channel of the BS
at which the message arrives (see below).

\medskip
\subsection{Beam splitter}

Here we construct a processing unit that acts as a BS, not by calculating the amplitudes according to QT,
but by processing individual events (see Fig.~\ref{BS}).
It consists of an input stage, a simple deterministic learning machine (DLM)~\cite{RAED05d,RAED05b,RAED05c,MICH05,ZHAO08b},
a transformation stage (T), an output stage (O) and
has two input and two output channels labeled by $k=0,1$.
We now define the operation of each stage explicitly.

\begin{itemize}
\item{Input stage: The DLM receives a message on either input channel $0$ or $1$, never on both channels simultaneously.
The arrival of a message on channel $0$ ($1$) is named a $0$ ($1$) event.
The input events are represented by the vectors ${\bf v}_l=(1,0)$ or ${\bf v}_l=(0,1)$ if the $l$th
event occurred on channel 0 or 1, respectively.
The DLM has six internal registers
${\bf Y}_{k,l}^{H}=(C_{k,l}^{H},S_{k,l}^{H}) ,$
${\bf Y}_{k,l}^{V}=(C_{k,l}^{V},S_{k,l}^{V}) ,$
${\bf Y}_{k,l}^{P}=(C_{k,l}^{P},S_{k,l}^{P}) $
and one internal vector ${\bf x}_{l}=( x_{0,l},x_{1,l}) $,
where $x_{0,l}+x_{1,l}=1$ and $x_{k,l}\geq 0$ for $k=0,1$ and all $l \geq 0$.
These seven two-dimensional vectors are labeled by the
message number $l$ to indicate that their values may change every time the DLM
receives a message. The DLM has storage for no more than fourteen numbers.

Upon receiving the $l$th input event, the DLM performs the following steps:
It stores the first two elements of message ${\bf y}_{k,l}$ in
its internal register
${\bf Y}_{k,l}^{H}=(C_{k,l}^{H},S_{k,l}^{H}) $,
the middle two elements of
${\bf y}_{k,l}$ in
${\bf Y}_{k,l}^{V}=( C_{k,l}^{V},S_{k,l}^{V})$,
and the last two elements of ${\bf y}_{k,l}$ in
${\bf Y}_{k,l}^{P}=( C_{k,l}^{P},S_{k,l}^{P})$.
Then, it updates its internal vector according to the rule
\begin{equation}
{\bf x}_{l}=\gamma {\bf x}_{l-1}+( 1-\gamma ) {\bf v}_l,
\label{eq_x}
\end{equation}
where $0<\gamma <1$.
Note that by construction $x_{0,l}+x_{1,l}=1$, $x_{0,l}\geq 0$ and  $x_{1,l}\geq 0$,
and the DLM stores information about the last message only.
The information carried by earlier messages is overwritten by
updating the internal registers.
From the solution of Eq.~(\ref{eq_x}),
\begin{equation}
{\bf x}_l=\gamma^l {\bf x}_{l-1}+( 1-\gamma )\sum_{j=0}^{l-1}\gamma^{l-j-1}{\bf v}_{j+1},
\end{equation}
the fact that in practice the sequence $\{{\bf v}_1, {\bf v}_2, \cdots, {\bf v}_K \}$ is finite,
and the usual trick to assume a periodic continuation of the sequence, we have
\begin{widetext}
\begin{eqnarray}
{\bf x}_{mK}&=&\gamma^K {\bf x}_{(m-1)K}+( 1-\gamma )\sum_{j=(m-1)K}^{mK-1}\gamma^{mK-j-1}{\bf v}_{j+1}
=\gamma^K {\bf x}_{(m-1)K}+( 1-\gamma )\sum_{j=0}^{K-1}\gamma^{K-j-1}{\bf v}_{j+1+(m-1)K}
\nonumber \\
&=&\gamma^K {\bf x}_{(m-1)K}+( 1-\gamma ){\bf f}_K
\label{eq_x2}
\end{eqnarray}
\end{widetext}
where
\begin{equation}
{\bf f}_K=\sum_{j=0}^{K-1}\gamma^{K-j-1}{\bf v}_{j+1},
\end{equation}
and $m\geq0$. From Eq.~(\ref{eq_x2}) we find
\begin{equation}
{\bf x}_{mK}=\gamma^{mK} {\bf x}_0+( 1-\gamma )\frac{1-\gamma^{mK}}{1-\gamma^K}{\bf f}_K,
\end{equation}
and hence
\begin{equation}
\lim_{m\rightarrow\infty}{\bf x}_{mK}=\frac{1-\gamma}{1-\gamma^K}\sum_{j=0}^{K-1}\gamma^{K-j-1}{\bf v}_{j+1},
\end{equation}
such that
\begin{equation}
\lim_{\gamma\rightarrow 1^-}\lim_{m\rightarrow\infty}{\bf x}_{mK}=\frac{1}{K}\sum_{j=0}^{K-1}{\bf v}_{j+1}.
\label{eq_x3}
\end{equation}
From Eq.~(\ref{eq_x3}), we conclude that as $\gamma\rightarrow 1^-$
the internal vector converges to the average of the vectors
${\bf v}_1, {\bf v}_2, \cdots, {\bf v}_K$ which represents
the relative frequency of input events at the two channels of the BS $(k=0,1)$.
The parameter $\gamma$ controls the speed of learning and also limits
the precision with which the internal vector can represent a sequence of constant input messages~\cite{RAED05d}.
Disregarding the fact that according to Eq.~(\ref{eq_x3}), we should let
$\gamma\rightarrow 1^-$ to obtain the limiting value of the average of the ${\mathbf v}$'s,
it is the only free parameter in the model. In practice, in the simulation we fix it once and for all.
}
\item{Transformation stage: The second stage (T) accepts the messages from the input stage, and transforms them
into a new eight-dimensional vector
\begin{equation}
{\bf T}=\frac{1}{\sqrt{2}}
\left(
\begin{array}{c}
C_{0,l}^{H}C_{0,l}^{P}\sqrt{x_{0,l}} - S_{1,l}^{H}C_{1,l}^{P}\sqrt{x_{1,l}} \\
C_{1,l}^{H}C_{1,l}^{P}\sqrt{x_{1,l}} + S_{0,l}^{H}C_{0,l}^{P}\sqrt{x_{0,l}} \\
C_{0,l}^{V}S_{0,l}^{P}\sqrt{x_{0,l}} - S_{1,l}^{V}S_{1,l}^{P}\sqrt{x_{1,l}} \\
C_{1,l}^{V}S_{1,l}^{P}\sqrt{x_{1,l}} + S_{0,l}^{V}S_{0,l}^{P}\sqrt{x_{0,l}} \\
C_{1,l}^{H}C_{1,l}^{P}\sqrt{x_{1,l}} - S_{0,l}^{H}C_{0,l}^{P}\sqrt{x_{0,l}} \\
C_{0,l}^{H}C_{0,l}^{P}\sqrt{x_{0,l}} + S_{1,l}^{H}C_{1,l}^{P}\sqrt{x_{1,l}} \\
C_{1,l}^{V}S_{1,l}^{P}\sqrt{x_{1,l}} - S_{0,l}^{V}S_{0,l}^{P}\sqrt{x_{0,l}} \\
C_{0,l}^{V}S_{0,l}^{P}\sqrt{x_{0,l}} + S_{1,l}^{V}S_{1,l}^{P}\sqrt{x_{1,l}} \\
\end{array}%
\right).
\label{T_vector}
\end{equation}%
If we rewrite the transformation $T$ using complex numbers,
we find that
\begin{equation}
\left( \begin{array}{c} 	b_0^H \\ b_0^V \\ b_1^H \\ b_1^V \end{array} \right)
=\frac{1}{\sqrt{2}}	\left( \begin{array}{cccc} 1 & 0 & i & 0 \\ 0 & 1 & 0 & i \\ i & 0 & 1 & 0 \\ 0 & i & 0 & 1 \end{array}
\right)\left( \begin{array}{c} 	a_0^H \\ a_0^V \\ a_1^H \\ a_1^V \end{array} \right),
\label{eqBS}
\end{equation}
which is the unitary transformation in the quantum theoretical description of a BS,
if $(a_0^H, a_0^V, a_1^H, a_1^V)$ and $(b_0^H, b_0^V, b_1^H, b_1^V)$
denote the input and output amplitudes of the photons with polarization
$H$ and $V$ in the $0$ and $1$ channels of a BS, respectively.
Note that in our simulation model there is no need to introduce
the (quantum theoretical) concept of a vacuum field, a requirement in the quantum optical description of a BS.
}
\item{Output stage: The final stage (O) sends out a messenger (representing a photon)
carrying the message
\begin{equation}
{\bf w}=\left(
\begin{array}{c}
w_{0,l}/s_{0,l} \\
w_{1,l}/s_{0,l} \\
w_{2,l}/s_{1,l} \\
w_{3,l}/s_{1,l} \\
s_{0,l}/s_{2,l} \\
s_{1,l}/s_{2,l}%
\end{array}%
\right) ,
\end{equation}%
where
\begin{eqnarray}
w_{0,l}&=&C_{0,l}^{H}C_{0,l}^{P}\sqrt{x_{0,l}} - S_{1,l}^{H}C_{1,l}^{P}\sqrt{x_{1,l}},\nonumber \\
w_{1,l}&=&C_{1,l}^{H}C_{1,l}^{P}\sqrt{x_{1,l}} + S_{0,l}^{H}C_{0,l}^{P}\sqrt{x_{0,l}},\nonumber \\
w_{2,l}&=&C_{0,l}^{V}S_{0,l}^{P}\sqrt{x_{0,l}} - S_{1,l}^{V}S_{1,l}^{P}\sqrt{x_{1,l}},\nonumber  \\
w_{3,l}&=&C_{1,l}^{V}S_{1,l}^{P}\sqrt{x_{1,l}} + S_{0,l}^{V}S_{0,l}^{P}\sqrt{x_{0,l}},\nonumber \\
s_{0,l}&=&\sqrt{w_{0,l}^{2}+w_{1,l}^{2}},\nonumber \\
s_{1,l}&=&\sqrt{w_{2,l}^{2}+w_{3,l}^{2}},\nonumber\\
s_{2,l}&=&\sqrt{w_{0,l}^{2}+w_{1,l}^{2}+w_{2,l}^{2}+w_{3,l}^{2}},\nonumber \\
\end{eqnarray}
through output channel 0 if $s_{2,l}^2>2r$ where
$0<r<1$ is a uniform pseudo-random number.
Otherwise, if $s_{2,l}^2 \le 2r$, the output stage sends
through output channel 1 the message%
\begin{equation}
{\bf z}=\left(
\begin{array}{c}
z_{0,l}/t_{0,l} \\
z_{1,l}/t_{0,l} \\
z_{2,l}/t_{1,l} \\
z_{3,l}/t_{1,l} \\
t_{0,l}/t_{2,l} \\
t_{1,l}/t_{2,l}
\end{array}
\right),
\end{equation}
where
\begin{eqnarray}
z_{0,l}&=&C_{1,l}^{H}C_{1,l}^{P}\sqrt{x_{1,l}} - S_{0,l}^{H}C_{0,l}^{P}\sqrt{x_{0,l}},\nonumber \\
z_{1,l}&=&C_{0,l}^{H}C_{0,l}^{P}\sqrt{x_{0,l}} + S_{1,l}^{H}C_{1,l}^{P}\sqrt{x_{1,l}},\nonumber \\
z_{2,l}&=&C_{1,l}^{V}S_{1,l}^{P}\sqrt{x_{1,l}} - S_{0,l}^{V}S_{0,l}^{P}\sqrt{x_{0,l}},\nonumber \\
z_{3,l}&=&C_{0,l}^{V}S_{0,l}^{P}\sqrt{x_{0,l}} + S_{1,l}^{V}S_{1,l}^{P}\sqrt{x_{1,l}},\nonumber \\
t_{0,l}&=&\sqrt{z_{0,l}^{2}+z_{1,l}^{2}},\nonumber \\
t_{1,l}&=&\sqrt{z_{2,l}^{2}+z_{3,l}^{2}},\nonumber \\
t_{2,l}&=&\sqrt{z_{0,l}^{2}+z_{1,l}^{2}+z_{2,l}^{2}+z_{3,l}^{2}}.\nonumber \\
\end{eqnarray}
}
\end{itemize}

The use of pseudo-random numbers to select the output channel
is not essential~\cite{RAED05b}. We use pseudo-random numbers to mimic the apparent unpredictability
of the experimental data only. Instead of a uniform pseudo-random number generator, any algorithm
that selects the output channel in a systematic manner might be employed as well~\cite{RAED05b}.
This will change the order in which messages are being processed but the content
of the messages will be left intact and the resulting averages do not change significantly.

\subsection{Polarizing Beam Splitter}

A polarizing beam splitter (PBS) is used to redirect the photons on the basis of their polarization ($H$ or $V$).
The structure of the event-based processor
that simulates a PBS is identical to the one of the BS and differs in the
details of the transformation stage only.
For the PBS, the transformation {\bf T} reads~\cite{ZHAO08b}
\begin{equation}
{\bf T}=
\left(
\begin{array}{c}
C_{0,l+1}^{H}C_{0,l+1}^{P}\sqrt{x_{0,l+1}} \\
S_{0,l+1}^{H}C_{0,l+1}^{P}\sqrt{x_{0,l+1}} \\
-S_{1,l+1}^{V}S_{1,l+1}^{P}\sqrt{x_{1,l+1}} \\
C_{1,l+1}^{V}S_{1,l+1}^{P}\sqrt{x_{1,l+1}} \\
C_{1,l+1}^{H}C_{1,l+1}^{P}\sqrt{x_{1,l+1}} \\
S_{1,l+1}^{H}C_{1,l+1}^{P}\sqrt{x_{1,l+1}} \\
-S_{0,l+1}^{V}S_{0,l+1}^{P}\sqrt{x_{0,l+1}} \\
C_{0,l+1}^{V}S_{0,l+1}^{P}\sqrt{x_{0,l+1}}
\end{array}%
\right) .
\end{equation}%

\subsection{Remaining optical components}

In contrast to the BS and PBS, in terms of message processing
the HWP and QWP are passive devices in the sense that the adaptive unit, the DLM,
is not required for a proper functioning of the devices.
As can be seen from the quantum theoretical description (see Appendix),
a HWP does not only change the polarization of the photon but
also changes its phase and a QWP additionally,
introduces a phase difference between the $H$ and $V$ components.
In our simulation model, the functionality of these optical components is implemented
through plane rotations of the vectors $(\cos\xi_{k,l},\sin\xi_{k,l})$,
$(\cos\psi^H_{k,l},\sin\psi^H_{k,l})$, and
$(\cos\psi^V_{k,l},\sin\psi^V_{k,l})$.

\subsection{Data gathering and analysis procedure}
In the simulation, the data is collected in the same manner as in the experiment.
Detector $D_0$ ($D_1$) registers the output events at channel 0 (1) (see Fig.~\ref{eraser}).
During a run of $N$ events, the algorithm generates the data set
\begin{equation}
\Gamma=\left\{x_{l}|l=1,...,N;\phi ; \theta_{HWP0} ; \theta_{HWP1}; \theta_{QWP}\right\} ,
\label{dataset}
\end{equation}
where $x_{l}=0,1$ indicates which detector fired ($D_{0}$ or $D_{1}$),
$\phi $ denotes the phase shift (proportional to the difference in time-of-flight of Path0 and Path1)
between the two interferometer arms and $\theta_{HWP0}$, $\theta_{HWP1}$, $\theta_{QWP}$ denote
the angles of the optical axis of the respective waveplates with the laboratory frame.
For fixed $\theta_{HWP0}$, $\theta_{HWP1}$, $\theta_{QWP}$ and $\phi$,
the number of detection events in detector 1 is given by $N_1=\sum_{l=1}^N x_l$
and $N_0=N-N_1$ is the number of detection events in detector 0.
The appearance of interference fringes is conveniently
characterized by the visibility~\cite{BORN64}
\begin{equation}
{\cal V} =\frac{N_{max}-N_{min}}{N_{max}+N_{min}},
\label{Visibility}
\end{equation}
where $N_{max}$ and $N_{min}$ denote
the maximum and minimum of $N_0$ for all $\phi\in[0,2\pi[$.
Notice that for the experiment depicted in Fig.~\ref{eraser},
the visibility is a function of $\theta_{HWP0}$, $\theta_{HWP1}$, and $\theta_{QWP}$.

\section{Simulation results}\label{sec6}
The processing units that simulate the optical components are connected in such a way that the simulation setup
is an exact one-to-one copy of the real experiment (see Fig.~\ref{eraser}).
The simulation procedure is as follows:
For each choice of $\phi$ in the range $[0,2\pi[$, we fix $\theta_{HWP0}$, $\theta_{HWP1}$ and $\theta_{QWP}$
and perform a simulation with $10^6$ events, randomly distributed over groups of $N_H =200$ or $N_V =200$ events
($\alpha_1 = H$ and $\alpha_2 = V$ in the notation of Section~\ref{sec4}).
Then for each choice of $\theta_{HWP0}$, $\theta_{HWP1}$, $\theta_{QWP}$, we repeat this procedure.
The result of these calculations form the data set $\Gamma$ (see Eq.~(\ref{dataset})).
From this data set, we compute the visibility according to Eq.~(\ref{Visibility}).
All simulations have been carried out with $\gamma=0.99$.

\subsection{Without QWP}

\setlength{\unitlength}{1cm}
\begin{figure}[t]
\begin{center}
\includegraphics[width=8cm]{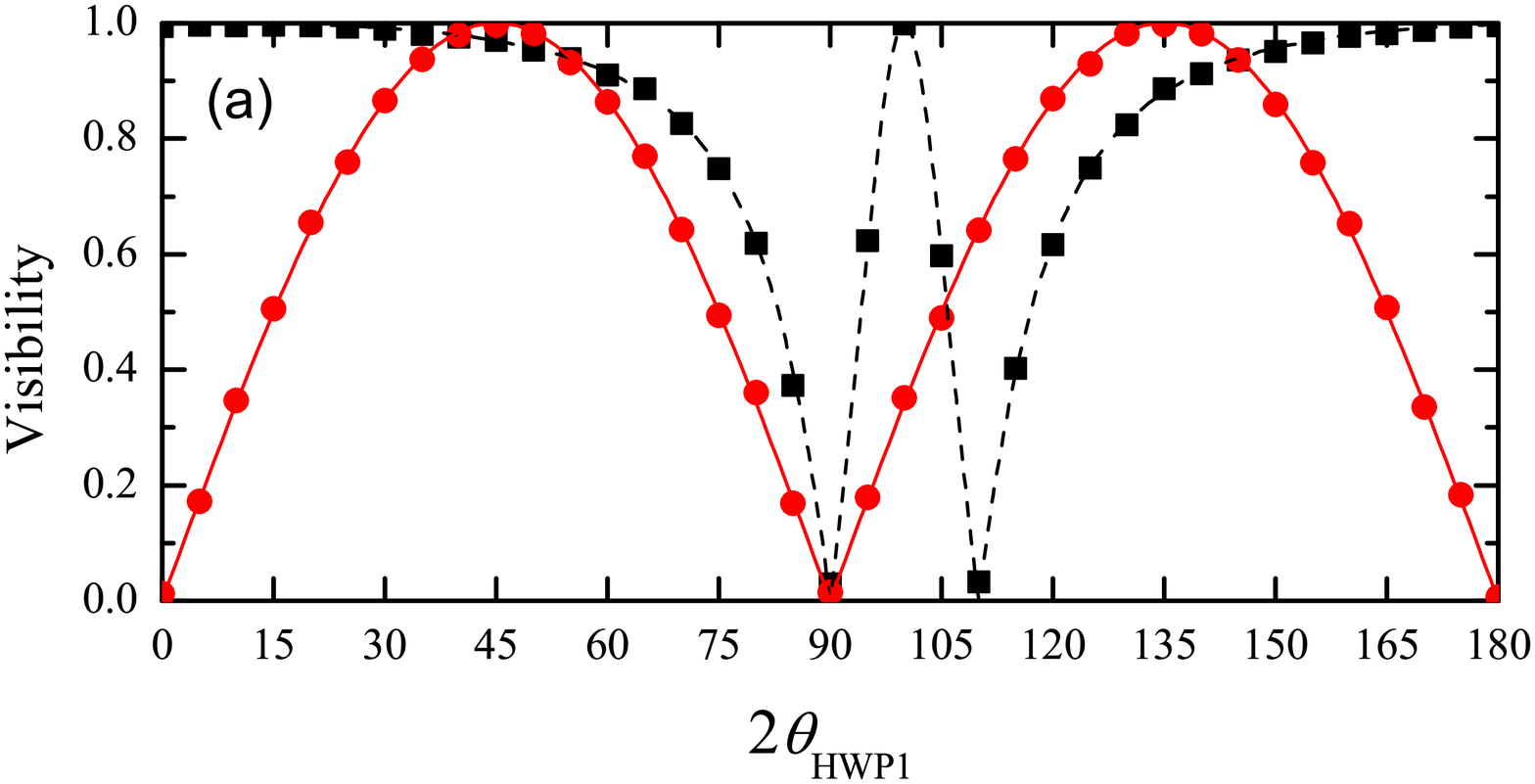}
\includegraphics[width=8cm]{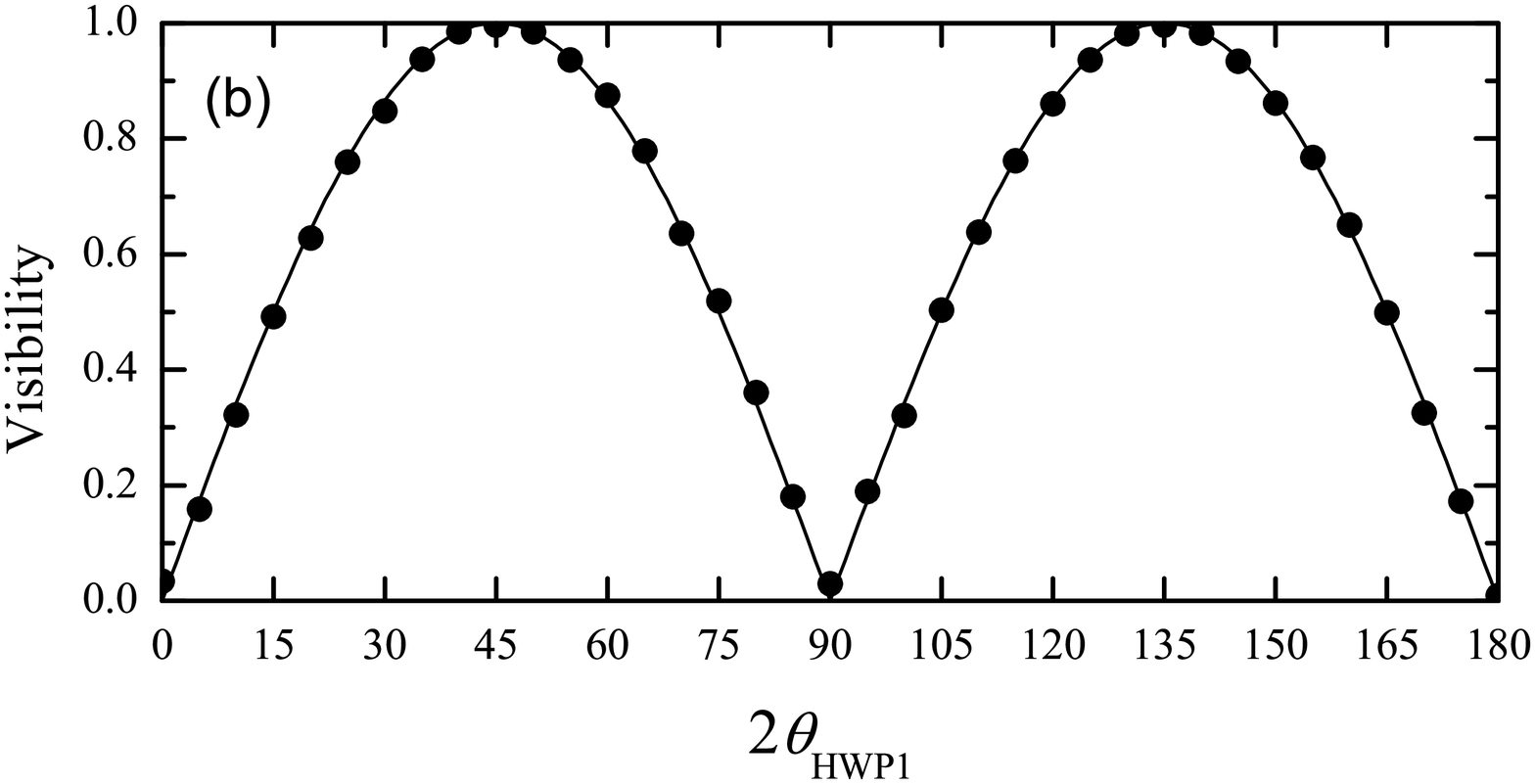}
\includegraphics[width=8cm]{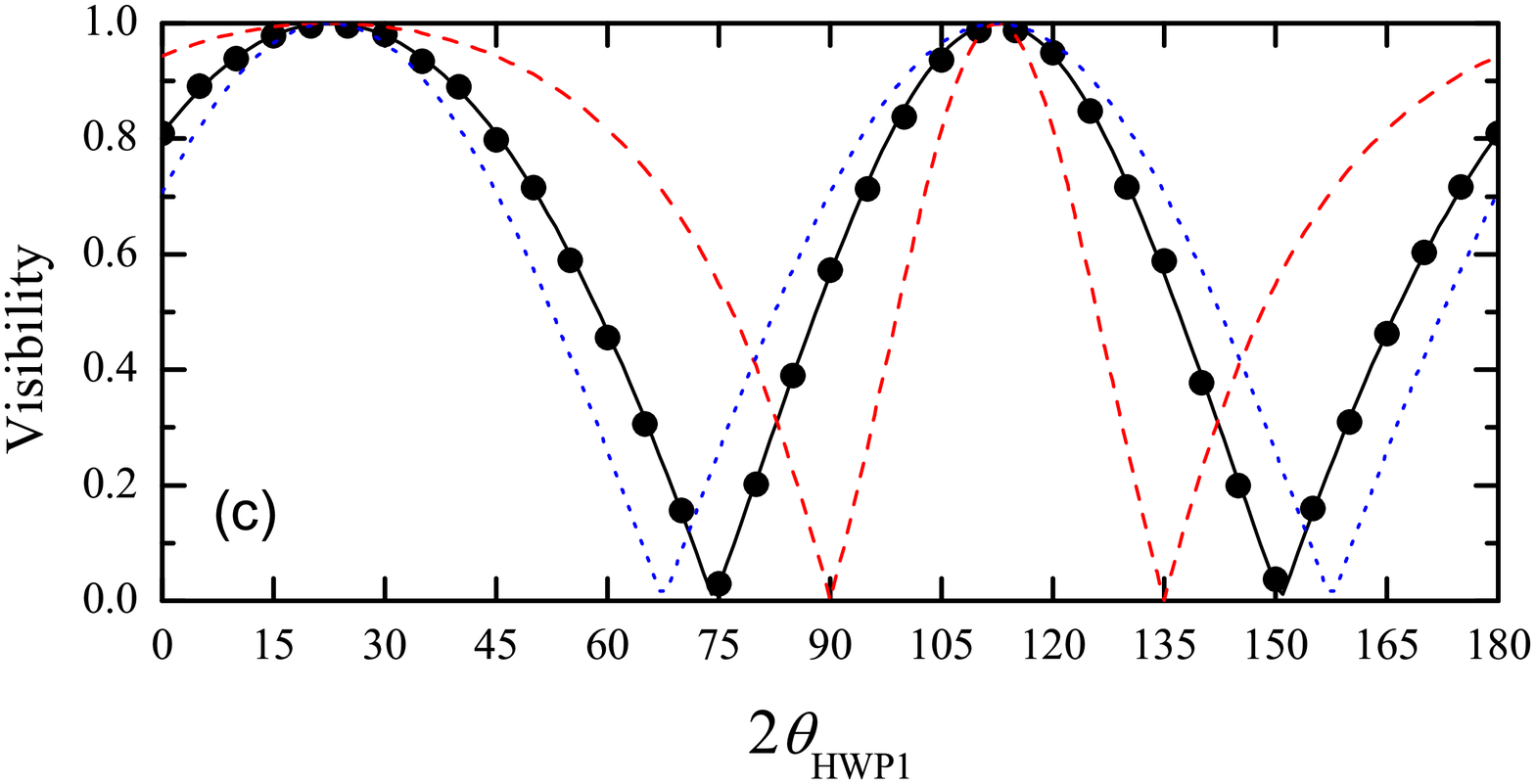}
\caption{Visibility as a function of the angle $2\theta_{HWP1}$ for the quantum eraser experiment
with the QWP removed (see Fig.~\ref{eraser}).
The markers (squares, bullets) and lines (solid, dashed) represent the event-by-event simulation data and
the quantum theoretical results (see Eqs.~(\ref{PUREV}) and (\ref{MIXED})), respectively.
(a) The source emits photons described by the pure vertically polarized state $V$ and $\theta_{HWP0}=45^\circ$
(red bullets and solid line), $\theta_{HWP0}=10^\circ$ (black squares and dashed line);
(b) The source emits photons described by the completely mixed state ($p_V=p_H=1/2$) and $\theta_{HWP0}=45^\circ$;
(c) The source emits photons described by a partially mixed state with $p_V=2/3$, $p_H=1/3$ and $\theta_{HWP0}=22.5^\circ$
(black bullets and solid line).
The red dashed and blue dotted curves represent the quantum theoretical results for
the pure vertically polarized state $V$ and the completely mixed state, respectively.
}
\label{results1}
\end{center}
\end{figure}

In Fig.~\ref{results1} we show our simulation results for the visibility as a function of $2\theta_{HWP1}$ for the quantum
eraser experiment with the QWP removed (see Fig.~\ref{eraser}).
First we consider the case in which the source emits photons that in QT are described by a pure, vertically polarized ($V$) state.
Each such photon, after passing through the first BS, has equal chance to end up in either of the two arms of the interferometer.
In our simulation, the messenger representing this photon carries the message
$\left( 0,0, \cos \psi _{0}^{V},\sin \psi _{0}^{V},0,1\right)$ (see Eq.~(\ref{message})).
If the photon follows Path0, it encounters HWP0, the optical axis of which makes an angle $\theta_{HWP0}$ with respect to the
laboratory frame. HWP0 rotates the polarization of the photon by an angle $2\theta_{HWP0}$~\cite{BORN64}.
The event-by-event simulation data and the results of QT are shown in Fig.~\ref{results1}(a).
The simulation data are in quantitative agreement with the averages calculated
from QT and in qualitative agreement with the experimental data (see Fig.~4(a) in Ref.~\cite{SCHW99}).

Next, we consider the case where in QT, the input to the quantum eraser is described by a (completely) mixed state.
In QT, a mixed state simply means that photons emitted by the source are described
by an incoherent mixture of horizontally and vertically polarized pure states.
In Section~\ref{sec4}, we explained how to implement mixed states in the event-based simulation approach.
The simulation data for a source emitting photons described by a (completely) mixed state are shown in Fig.~\ref{results1}(b) and (c).
Also in this case, our simulation data are in quantitative agreement with the averages computed from QT and in qualitative agreement
with the experimental results reported in Ref.~\cite{SCHW99} (see Fig.~4(b) and (c)).

\subsection{With QWP}

\setlength{\unitlength}{1cm}
\begin{figure}[t]
\begin{center}
\includegraphics[width=8cm]{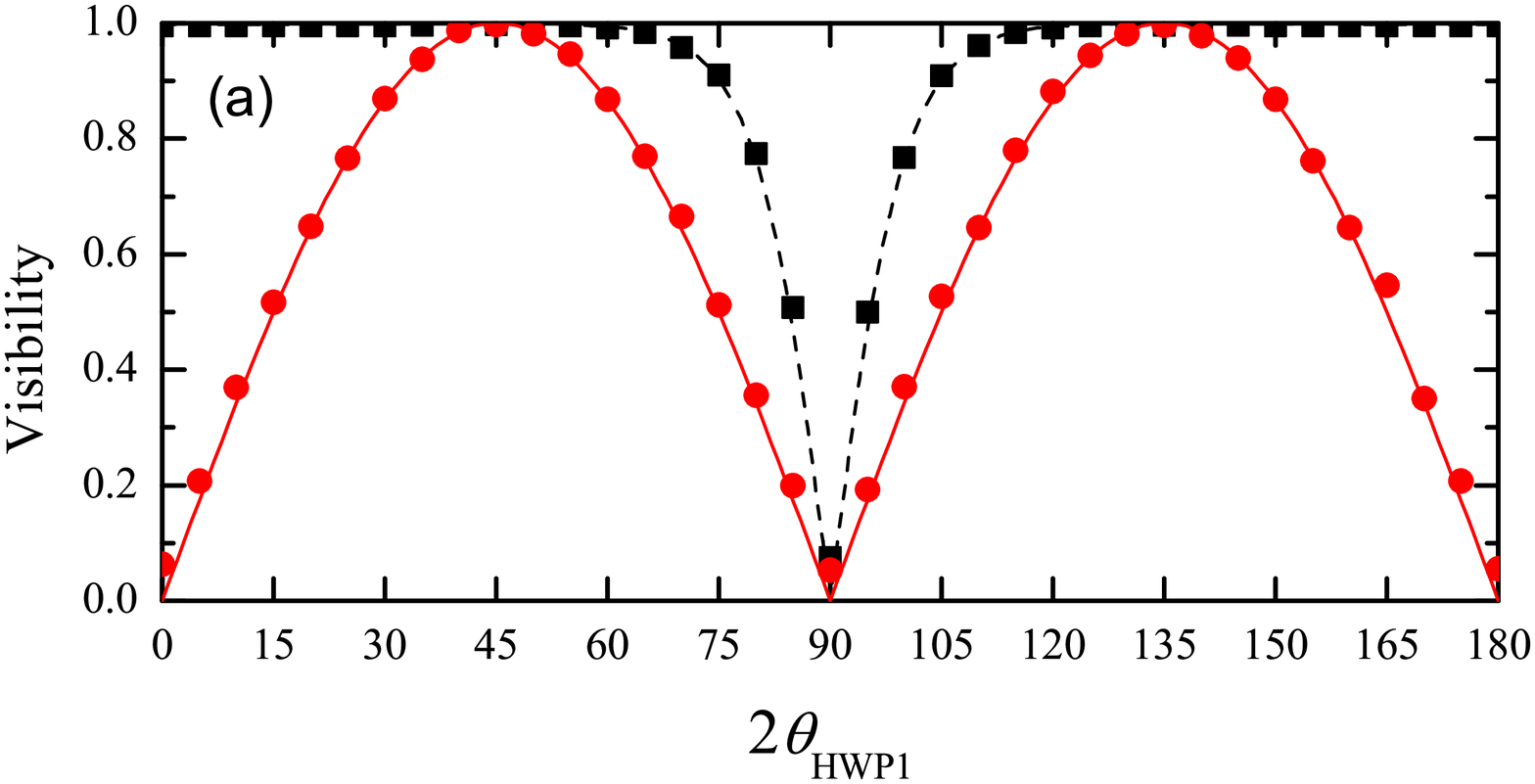}
\includegraphics[width=8cm]{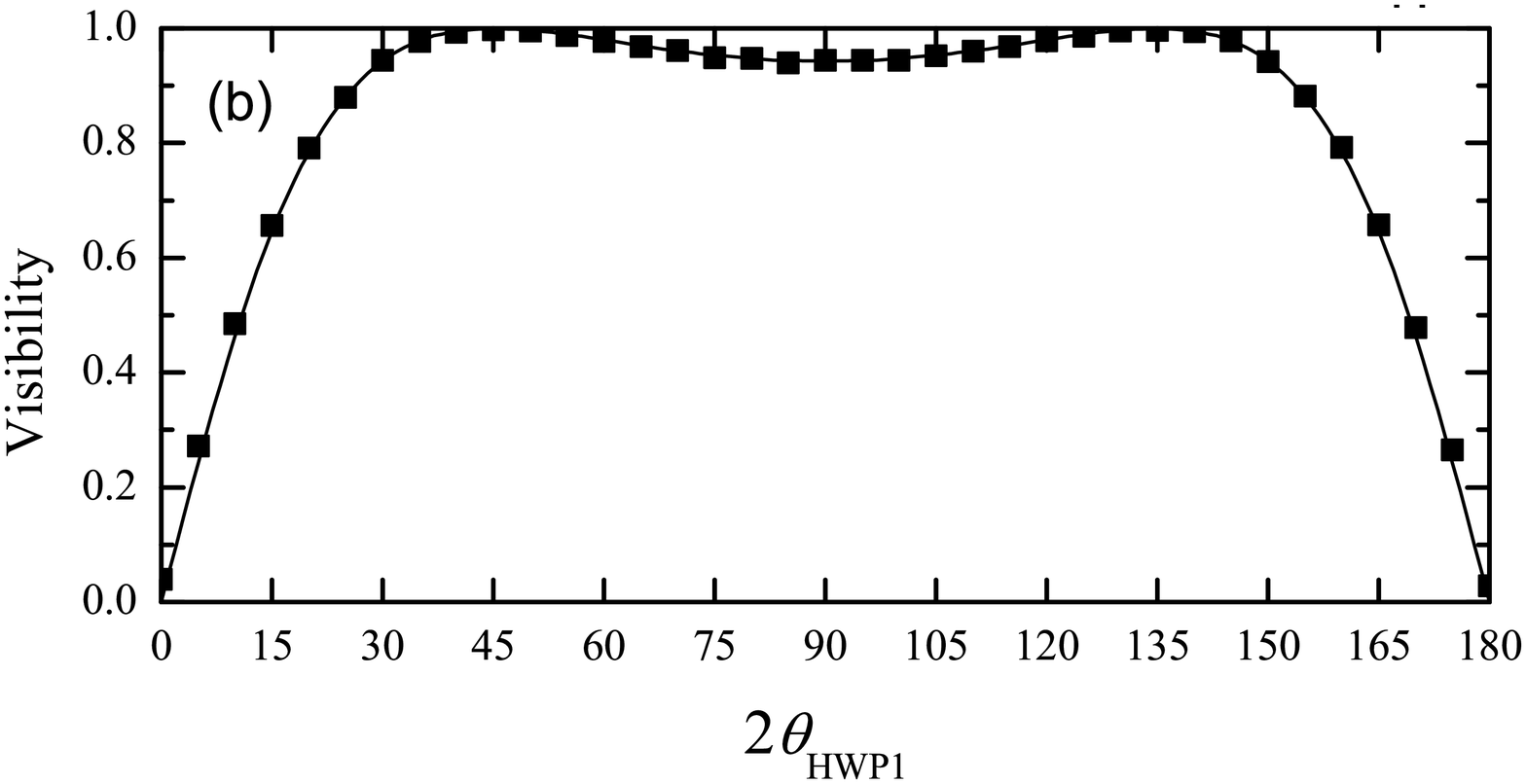}
\caption{Visibility as a function of the angle $2\theta_{HWP1}$ for the quantum eraser experiment depicted in Fig.~\ref{eraser} with $\theta_{QWP}=0$.
The markers (squares, bullets) and lines (solid, dashed) represent the event-by-event simulation data and
the quantum theoretical results (see Eq.~(\ref{PUREXI})), respectively.
(a) The source emits photons described by the pure vertically polarized state $V$ and $\theta_{HWP0}=45^\circ$
(red bullets and solid line), $\theta_{HWP0}=10^\circ$ (black squares and dashed line);
(b) The source emits photons described by the pure $\xi=45^\circ$-polarized state and $\theta_{HWP0}=22.5^\circ$.
}
\label{results2}
\end{center}
\end{figure}

In Fig.~\ref{results2} we present some simulation data for the case that the QWP is present, see Fig.~\ref{eraser},
and $\theta_{QWP}=0$. If $\theta_{QWP}=0$, the QWP does not change the polarization of the photons
but changes their phase.
We only consider the case that the single-photon source emits photons that in QT, are described by a pure state.
Figure~\ref{results2}(a) shows simulation data corresponding to incoming $V$-polarized photons,
for $\theta_{HWP0}=45^\circ$ (red bullets) and $\theta_{HWP0}=10^\circ$ (black squares).
In Fig.~\ref{results2}(b) we show the simulation data for
the source that emits photons in a state that QT would characterize with $\xi=45^\circ$,
and $\theta_{HWP0}=22.5^\circ$.
In our simulation, this state is represented by messengers that carry the message
$(\cos \psi _{0}^{H},\sin \psi _{0}^{H}, \cos \psi _{0}^{V},\sin \psi _{0}^{V},1/\sqrt{2},1/\sqrt{2}) $.
As in all other cases shown, the agreement between the event-based simulation data and QT is excellent.

\section{Discussion}\label{sec7}

We have demonstrated that our classical, locally causal, particle-like simulation
approach reproduces the results of the quantum eraser experiment~\cite{SCHW99} and the results of quantum theory describing the averages of
these experimental results.

During the event-by-event simulation of the quantum eraser experiment we always have full which-way information of the photons (messengers)
since we can always track them.
Nevertheless, depending on the settings of the optical apparatuses, the photons build up an interference pattern at the detector.
Although the appearance of an interference pattern is commonly considered to be characteristic for a wave, we have demonstrated that,
as in experiment, it can also be built up by many photons. These photons have full which-way information and arrive one-by-one at a detector.
Hence, even in the case that the source emits single photons, described by a pure state in quantum theory, and that ${\cal V}=1$, commonly
associated with full wave character, the photons in our simulation model have full which-way information.
A consequence of our model is thus that the relation ${\cal V}^2+{\cal D}^2\leq 1$ cannot be regarded as quantifying the notion of
complementarity:
Our model always allows a particle-only description of the quantum eraser experiment,
independent of the purity of the state describing the photons in quantum theory.

In summary, concepts of quantum theory applied to individual events
fail to provide a logically consistent explanation for the experimental observation of single detector ``clicks''
building up an interference pattern
and leave no option but to postulate that ``this is the way it is''.
In contrast, our event-based simulation model, a classical locally causal dynamical system,
reproduces the results of quantum theory without making reference to the solution
of a wave equation and provides a simple, particle-based mental picture for what each
individual photon experiences as it travels from the source to the detector.
Just like in the experiments, our model produces data sets Eq.~(\ref{dataset}) which can be given to a third
party for analysis long after the simulation has been finished.
Because of the strong similarity between the experimental
and simulation data sets the third party will
have a very hard time, if possible at all, to identify the
data sets as originating from a so-called ``quantum experiment'' or from a ``classical simulation model''.

Finally, we would like to emphasize that the algorithms used to simulate
the optical components of the quantum eraser have not designed to exclusively simulate
this particular example but they can be used to reproduce the results of many
other quantum optics experiments 
as well~\cite{RAED05d,RAED05b,RAED05c,MICH05,RAED06c,RAED07a,RAED07b,RAED07c,ZHAO07b,ZHAO08,ZHAO08b,JIN09a,JIN09b}.

\medskip
\appendix
\section*{Appendix}
According to quantum theory (QT), photons in a pure state are described by the state vector
\begin{equation}
|\alpha\rangle=
\left( \begin{array}{c} 	a_0^H \\ a_0^V \\ a_1^H \\ a_1^V \end{array} \right),
\label{app0}
\end{equation}
where $H$ and $V$ refer to the horizontal and vertical direction of polarization
and the subscripts refer to the wave in Path0 and Path1, respectively.
Within QT, the action of the various optical components is defined by the matrices

\begin{eqnarray}
T_{BS}&=&\frac{1}{\sqrt{2}}	\left( \begin{array}{cccc} 1 & 0 & i & 0 \\ 0 & 1 & 0 & i \\ i & 0 & 1 & 0 \\ 0 & i & 0 & 1 \end{array}\right)
,
\end{eqnarray}

\begin{eqnarray}
T_{PBS}&=&{\left( \begin{array}{cccc} 1 & 0 & 0 & 0 \\ 0 & 0 & 0 & i \\ 0 & 0 & 1 & 0 \\ 0 & i & 0 & 0 \end{array} \right)},
\end{eqnarray}

\begin{eqnarray}
T_{HWP0}(\theta)&=&-i\left(\begin{array}{cccc}
\phantom{-}c & \phantom{-}s& \phantom{-}0 &\phantom{-}0 \\
\phantom{-}s & -c &\phantom{-} 0& \phantom{-}0\\
\phantom{-}0&\phantom{-}0 & \phantom{-}1 &\phantom{-} 0 \\
\phantom{-}0&\phantom{-}0 &\phantom{-} 0 &\phantom{-} 1
\end{array}\right)
,
\end{eqnarray}
\begin{eqnarray}
T_{HWP1}(\theta)&=&-i\left(\begin{array}{cccc}
\phantom{-}1 & \phantom{-}0& \phantom{-}0 &\phantom{-}0 \\
\phantom{-}0 & \phantom{-}1&\phantom{-} 0& \phantom{-}0\\
\phantom{-}0&\phantom{-}0 & \phantom{-}c &\phantom{-} s \\
\phantom{-}0&\phantom{-}0 &\phantom{-} s & -c
\end{array}\right)
,
\end{eqnarray}
\begin{eqnarray}
T_{QWP}(\theta)&=&\frac{1}{\sqrt{2}}\left(
\begin{array}{cccc}
\phantom{-}1 & \phantom{-}0&0 &0 \\
\phantom{-}0 & \phantom{-}1& 0&0\\
\phantom{-}0 &\phantom{-}0 &1 - i c & - i s \\
\phantom{-}0 &\phantom{-}0 &- i s   & 1 + i c
\end{array}\right),
\end{eqnarray}
where $\theta$ denotes the angle of the optical axis with respect to the laboratory frame, $c=\cos2\theta$
and  $s=\sin2\theta$.

\begin{widetext}
Using these expressions, it is somewhat tedious but straightforward to calculate
the visibility Eq.~(\ref{Visibility}). We list the expressions for the cases for which we perform
event-based simulations.

\begin{enumerate}
\item{With the QWP removed, see Fig.~\ref{eraser},
and for incoming photons that are described by a pure state of polarization $\xi$:
\begin{equation}
{\cal V}=\left| \frac{2\sin (\xi-2\theta_0+2\theta_1) \sin (\xi-2\theta_1) }
{\sin^2 (\xi-2\theta_0+2\theta_1) + \sin^2 (\xi-2\theta_1)} \right|
,
\label{PUREV}
\end{equation}
where $\theta_0=\theta_{HWP0}$ and $\theta_1=\theta_{HWP1}$.
}
\item{With the QWP removed and for incoming photons described by a mixed-state photon input with
$p_V/p_H=\tan^2\beta$:
\begin{equation}
V=\left| \frac{2\sin (2\theta_0-2\theta_1) \sin (2\theta_1) + 2\tan^2\beta\cos(2\theta_0-2\theta_1) \cos (2\theta_1) }
{\sin^2 (2\theta_0-2\theta_1) + \sin^2 (2\theta_1)+\tan^2\beta[\cos^2(2\theta_0-2\theta_1)+\cos^2 (2\theta_1)]} \right|
.
\label{MIXED}
\end{equation}
}
\item {With the QWP present, $\theta_{QWP}=0^\circ$,
and for incoming photons that are described by a pure state of polarization $\xi$:
\begin{equation}
V=\left| \frac{\left[\sin^2 4\theta_1 \sin^2 (2\xi-2\theta_0) + 4[\cos^2 2\theta_1 \sin(\xi-2\theta_0)\sin\xi-\sin^2 2\theta_1 \cos(\xi-2\theta_0)\cos\xi]^2 \right]^{1/2}}
{\cos^2 2\theta_1 \sin^2 (\xi-2\theta_0) + \sin^2 2\theta_1 \cos^2 (\xi-2\theta_0) + \sin^2 2\theta_1 \cos^2 \xi
+ \cos^2 2\theta_1 \sin^2 \xi} \right| .
\label{PUREXI}
\end{equation}
}
\end{enumerate}
\end{widetext}

\bibliography{../../epr}

\end{document}